\documentclass[prd,preprint,eqsecnum,nofootinbib,amsmath,amssymb,
               tightenlines,dvips]{revtex4}
\usepackage{graphicx}
\usepackage{bm}%	bold math
\input epsf

\begin {document}

%%%%%%%%%%%%%%%%%%%%%%%%%%%%%%%%%%%%%%%%%%%%%%%%%%%%%%%%%%%%%%%%%%%%%%%%%%%%%%%

%\def\gsim{\mbox{~{\raisebox{0.4ex}{$>$}}\hspace{-1.1em}
%	{\raisebox{-0.6ex}{$\sim$}}~}}
%\def\lsim{\mbox{~{\raisebox{0.4ex}{$<$}}\hspace{-1.1em}
%	{\raisebox{-0.6ex}{$\sim$}}~}}

\def\ca{C_{\rm A}}
\def\cs{C_s}
\def\cf{C_{\rm F}}

\def\df{d_{\rm F}}
\def\da{d_{\rm A}}
\def\h{\bm h}

\def\qperp{{\bm q}_\bot}

\def\md{m_{\rm D}}

\def\half{\tfrac{1}{2}}

\def\Real{{\rm Re}}

\def\alphas{\alpha_{\rm s}}
\def\p{{\bm p}}

\def\QTo{\hat Q_{\perp0}}
\def\gammaE{\gamma_{\rm E}^{}}
\def\Nf{N_{\rm f}}

\def\q{{\bm q}}

\def\ta{t_{\rm A}}
\def\tf{t_{\rm F}}

\def\r{{\bm r}}

\def\qhat{\hat{\bar q}}
\def\num{\Xi}

%%%%%%%%%%%%%%%%%%%%%%%%%%%%%%%%%%%%%%%%%%%%%%%%%%%%%%%%%%%%%%%%%%%%%%%%%%%%%%%

%\preprint {UW/PT 03--01}

\title
    {
     High-energy jet quenching in weakly-coupled quark-gluon plasmas
    }

\author{Peter Arnold and Wei Xiao}
\affiliation
    {%
    Department of Physics,
    University of Virginia, Box 400714,
    Charlottesville, Virginia 22904, USA
    }%

\date {\today}

\begin {abstract}%
{%
$\hat q$ is the average squared transverse momentum transfer
per unit length to a high-energy particle traversing a
QCD medium such as a quark-gluon plasma.  We find the
(UV-regulated) value of $\hat q$ to leading order in the
weak coupling limit, $\alphas(T) \ll 1$.
We then use this value to generalize previous analytic
results on the gluon bremsstrahlung and pair production rates
for massless high-energy particles in a weakly-coupled quark-gluon plasma,
at next-to-leading logarithmic order.
}%
\end {abstract}

\maketitle
\thispagestyle {empty}

%%%%%%%%%%%%%%%%%%%%%%%%%%%%%%%%%%%%%%%%%%%%%%%%%%%%%%%%%%%%%%%%%%%%%%%%%%%%%%%

\section {Introduction}
\label{sec:intro}

In a weakly-coupled quark-gluon plasma,
energy loss of high-energy jets is dominated
by gluon bremsstrahlung and pair production.  One of the elements
that goes into the calculation of gluon bremsstrahlung rates in
the high-energy limit is the parameter $\hat q$ --- the average
squared transverse momentum transfer per unit length to a
high-energy particle due to elastic collisions with plasma
particles.  In this paper, we find the (UV-regulated) value of
$\hat q$ to leading order in the weak coupling limit.  We then use
it to generalize previous analytic results on the gluon
bremsstrahlung and pair production rates at leading order in
coupling and next-to-leading order in inverse powers of
$\ln(E/T)$, where $E$ is the energy of a jet particle.

% ---------------------------------------------------------------------------

\subsection {The parameter $\hat q$}

We will define the parameter $\hat q$ by
\begin {equation}
   \hat q(\Lambda) \equiv
   \int_{q_\perp < \Lambda} d^2 q_\perp \>
   \frac{d\Gamma_{\rm el}}{d^2q_\perp}
   \, q_\perp^2 ,
\label {eq:qhatdef}
\end {equation}
where $\Gamma_{\rm el}$ is the rate for elastic collisions with plasma
particles, $q_\perp$ is the transverse momentum transfer in a single
such collision, and
$\Lambda$ is an ultraviolet cut-off whose physical scale will be set
later when we discuss gluon bremsstrahlung.
The value of $\hat q$ is proportional to a factor
of the quadratic Casimir $C_R$ for the color representation $R$ of
the high-energy particle.  It will be convenient to factor out
this dependence by defining $\qhat$ by
\begin {equation}
   \hat q = C_R \qhat .
\end {equation}

The differential elastic cross-section has the known limiting
forms%
\footnote{
   The simple formula for the small $q_\perp$ case is taken from
   Aurenche, Gelis, and Zakaret \cite{AGZ}.
   The $d\Gamma_{\rm el}/d^2q_\perp$ presented here corresponds
   to the notation $g^2 C_R \ {\cal A}(q_\perp) / (2\pi)^2$ of
   Refs. \cite{AMY3,ArnoldDogan}.  See Appendix A of
   Ref.\ \cite{timelpm1}.
}
\begin {equation}
   \frac{d\Gamma_{\rm el}}{d^2q_\perp}
   \simeq
   \frac{C_R}{(2\pi)^2} \times
   \begin {cases}
      \displaystyle\frac{g^2 T \md^2}{q_\perp^2 (q_\perp^2+\md^2)} \,,
      & q_\perp \ll T, \\[15pt]
      \displaystyle\frac{g^4 {\cal N}}{q_\perp^4} \,,
      & q_\perp \gg T,
   \end {cases}
\label {eq:dsig}
\end {equation}
where
$\md$ is the Debye mass and ${\cal N}$ is the density of plasma
particles times color group factors (details below).

If we wanted $\qhat(\Lambda)$ for $\Lambda \ll T$, we could simply
use the first case of (\ref{eq:dsig}) to obtain%
\footnote{
   For (\ref{eq:qhatsmall}), see also Eq.\ (13) of Ref.\ \cite{BMT}
   and the relation to
   Ref.\ \cite{MooreTeaney} discussed after Eq. (61) of Ref.\ \cite{BMT}.
}
\begin {equation}
   \qhat(\Lambda) \simeq
   \int_{q_\perp < \Lambda}
      \frac{d^2 q_\perp}{(2\pi)^2} \> \frac{g^2 T \md^2}{q_\perp^2+\md^2}
   \simeq
   2 \alpha T \md^2 \ln\left(\frac{\Lambda}{\md}\right)
   \qquad
   (\md \ll \Lambda \ll T) .
\label {eq:qhatsmall}
\end {equation}
However, as we shall discuss momentarily, bremsstrahlung for
sufficiently high energy particles depends on $\qhat(\Lambda)$ for
$\Lambda \gg T$.  For this, we will need to find
$d\Gamma_{\rm el}/d^2 q_\perp$ for $q_\perp \sim T$ as well.
Our final result for $\qhat(\Lambda)$ in this case is given in
Sec.\ \ref{sec:results} below, and the derivation follows in
Sec.\ \ref{sec:qhat}.

Finding $\qhat(\lambda)$ for $\Lambda \gg T$ is an interesting problem
of principle: It's always good to understand how to get precise answers
to questions in the limit of weak coupling.  But it's important to note
that, even within the context of the weak coupling limit, getting it
right will not have a large effect on the answer.  As noted in
Ref.\  \cite{ArnoldDogan}, the numerators $g^2 \md^2 T$ and
$g^4 {\cal N}$ in (\ref{eq:dsig}) differ by only about 15\% for
3-flavor QCD.  So one would not make a large error by using the
simple formula (\ref{eq:qhatsmall}) even when $\Lambda \gg T$.
Nonetheless, in the present work we aim to find the exact
weak-coupling answer for this limit.

% ---------------------------------------------------------------------------

\subsection {Gluon bremsstrahlung}

When very high-energy particles travel through a weakly-coupled
quark-gluon plasma, the dominant energy loss mechanism is through hard
bremsstrahlung or pair creation.  Each time a high energy particle
collides with a plasma particle, there is the potential for
bremsstrahlung or pair creation.  At high energies (parametrically $E
\gtrsim T$), the quantum mechanical duration (formation time)
of a splitting
process becomes larger than the mean free time between the underlying
collisions with plasma particles, leading to coherence effects that
reduce the bremsstrahlung or pair production rate.  This is known as the
Landau-Pomeranchuk-Migdal (LPM) effect.  In this paper, we consider
the calculation of such splitting rates for particles with large energies
$E \gg T$, where $T$ is the temperature of the plasma.  We will consider
the case of a time-independent, uniform, infinite quark-gluon plasma.
In practice, that means that the plasma temperature should not vary
significantly over the time and distance scales associated with the
formation time of the splitting process.

A calculation of these rates to leading-order in $\alphas$ was
performed by Jeon and Moore \cite{JeonMoore}, based on the formalism
of Arnold, Moore, and Yaffe (AMY) \cite{AMY1,AMY2,AMY3}.  These calculations
require numerical solutions of integral equations.
It was noted earlier by Baier, Dokshitzer, Mueller, Peigne and Schiff
(BDMPS) \cite{BDMPS1,BDMPS2,BDMPS3,BDMS},
however, that analytic solutions can be found if the
logarithm of the energy is large, and they performed a leading-log
analysis.  Arnold and Dogan \cite{ArnoldDogan} recently extended
this analysis to next-to-leading order in inverse powers of
$\ln(E/T)$.  The results of all of these analysis depend on
the differential rate $d\Gamma_{\rm el}/d^2 q_\perp$ for elastic
scattering of the high energy particle off of a plasma particle,
as a function of the transverse momentum exchange $q_\perp$.
The full numerical calculations of rates to leading order in
$\alphas$ by Jeon and Moore \cite{JeonMoore}, and its approximation
to next-to-leading log (NLL) by Arnold and Dogan \cite{ArnoldDogan},
were carried out using the $q_\perp \ll T$ case of (\ref{eq:dsig}).
However, the relevant range of $q_\perp$ which contributes to the
calculation grows as the energy $E$ of the splitting parton grows.
Roughly speaking, as the formation time gets longer, that time subsumes
more and more elastic collisions, which means that the total momentum
transfer $Q_\perp$ from the plasma to the splitting particle during
the formation time grows larger.
$Q_\perp$ grows parametrically as%
\footnote{
   Eq. (\ref{eq:Qperp}) is for the case where the gluon momentum
   fraction $x \sim 1$ and $1-x \sim 1$, which dominates energy loss.
   More generally, the scale of
   the formation time is determined by replacing $E$ by the smallest
   of the final particle energy and bremsstrahlung gluon energy.
   The estimate (\ref{eq:Qperp}) can be understood as
   $Q_\perp^2 \sim \hat q t_{\rm form}$, where $t_{\rm form}$ is
   the formation time, combined with the standard
   result that $t_{\rm form} \sim \sqrt{E/\hat q}$ for
   $x \sim 1-x \sim 1$.  See, for example, Sec. 3 of Ref.\
   \cite{BDMPS2}, and Ref.\ \cite{baier}.
}
\begin {equation}
   Q_\perp \sim (\hat q E)^{1/4} .
\label {eq:Qperp}
\end {equation}
The total momentum transfer $Q_\perp$
is made up of many individual momentum transfers $q_\perp$,
which range in scale from $\md$ to $Q_\perp$ itself.
At high enough energy, the assumption $q_\perp \ll T$ therefore
breaks down for the upper end of this range.
Based on (\ref{eq:Qperp}), the condition $Q_\perp \ll T$ for assuming
all $q_\perp \ll T$ is
\begin {equation}
   E \ll \frac{T^4}{\hat q} \,.
\end {equation}
Using (\ref{eq:qhatsmall}) and $\md\sim gT$, this condition is parametrically
\cite{ArnoldDogan}
\begin {equation}
  E \ll \frac{T}{g^4\ln(1/g)} \,.
\label {eq:Esmall}
\end {equation}
In the current work, we will redo Arnold and Dogan's
next-to-leading log calculation of hard bremsstrahlung and pair
production, rewriting the answer in terms of $\qhat$.
We will then be able to use our result for $\qhat(\Lambda)$ with
$\Lambda \gg T$ to generalize the previous results to the new case
\begin {equation}
  E \gg \frac{T}{g^4\ln(1/g)} \,.
\label {eq:Elarge}
\end {equation}
Our result is given in Sec.\ \ref{sec:results} below,
and the few modifications
to the original derivation of Arnold and Dogan \cite{ArnoldDogan}
are outlined in Appendix \ref{app:NLL}.

% ---------------------------------------------------------------------------

\subsection {Weak coupling and \boldmath$\md \ll T$}

In this paper, we ruthlessly work in the weak coupling limit.
In particular, the Debye mass $\md$ is parametrically of order $gT$,
and so we shall formally assume that $\md \ll T$.  In practice, however,
one interest in weak coupling calculations is to see what they give if
optimistically applied to realistic situations where the coupling is not
terribly small.  Unfortunately, the weak coupling result for
$\md$ for massless 3-flavor QCD gives, for example,
$\md \simeq 2.4 T$ at $\alphas \simeq 0.3$.  Exactly how terrible it is
to treat $\md \ll T$ depends on the details of the calculation.
In this paper, we simply explore the weak coupling
limit and will not consider possible improvements that might
be made by not assuming $\md \ll T$.
Recent progress on going beyond this approximation has been
made by Caron-Huot \cite{simon}, who pushes the
calculation of $d\Gamma_{\rm el}/d^2q_\perp$ and
$\hat q$ to next-to-leading order in coupling $g$.

%============================================================================

\section {Results}
\label {sec:results}

\subsection {Notation}

Throughout, we use the same notation for group factors as in
Ref.\ \cite{ArnoldDogan}.  For a color representation $R$,
$C_R$ is the quadratic Casimir, $d_R$ is the dimension,
and $t_R = d_R C_R/d_A$ is the trace normalization.  For QCD,
\begin {equation}
   \ca = 3,
   \qquad
   \cf = \tfrac43 \,,
   \qquad
   \da = 8,
   \qquad
   \df = 3,
   \qquad
   \ta = 3,
   \qquad
   \tf = \tfrac12
\end {equation}
for gluons (A) and quarks (F).

We will use $\num_{\rm b}$ and $\num_{\rm f}$ to represent the
number of spin and flavor degrees of freedom in plasma bosons and
fermions, respectively, times the corresponding trace normalizations
$t_{\rm R}$.
In $\Nf$-flavor QCD,
\begin {equation}
   \num_{\rm b} = 2\ta = 6,
   \qquad
   \num_{\rm f} = 4\Nf\tf = 2 \Nf .
\end {equation}

The Debye mass and the weighted density ${\cal N}$ appearing
in (\ref{eq:dsig}) are given by the following formulas, which are
cast in a form that will be useful later on:
\begin {subequations}
\label {eq:mdN}
\begin {equation}
   \md^2 = \bigl[\num_{\rm b} \zeta_+(2) + \num_{\rm f} \zeta_-(2)\bigr]
           \, \frac{g^2 T^2}{\pi^2}
   = (1 + \tfrac16 \Nf ) g^2 T^2
\label {eq:md}
\end {equation}
and
\begin {equation}
   {\cal N} = \bigl[\num_{\rm b} \zeta_+(3) + \num_{\rm f} \zeta_-(3)\bigr]
           \, \frac{T^3}{\pi^2}
   = \frac{\zeta(3)}{\zeta(2)} \, (1 + \tfrac14\Nf) T^3 .
\end {equation}
\end {subequations}
Here the functions $\zeta_{\pm}(z)$ are the bosonic and fermionic versions
of the Riemann $\zeta$ function,
\begin {equation}
   \zeta_{\pm}(s) \equiv \sum_{k=1}^\infty \frac{(\pm)^{k-1}}{k^s} ,
\end {equation}
and so
\begin {equation}
   \zeta_+(s) = \zeta(s) ,
   \qquad
   \zeta_-(s) = (1-2^{1-s}) \, \zeta(s) .
\end {equation}
Recall that $\zeta(2) = \pi^2/6$.

% ---------------------------------------------------------------------------

\subsection{Result for \boldmath $\qhat$}

Our final result for $\qhat(\Lambda)$ for large cut-off $\Lambda$ is
\begin {subequations}
\label {eq:qhateqs}
\begin {equation}
   \qhat(\Lambda) =
   \bigl[\num_{\rm b} \, {\cal I}_+(\Lambda)
        +\num_{\rm f} \, {\cal I}_-(\Lambda)
   \bigr]
   \frac{g^4 T^3}{\pi^2}
   \qquad
   (\Lambda \gg T)
\label {eq:qhatlarge}
\end {equation}
with
\begin {equation}
   {\cal I}_\pm(\Lambda) \simeq
   \frac{\zeta_\pm(3)}{2\pi} \ln\left(\frac{\Lambda}{\md}\right)
   + \Delta {\cal I}_\pm ,
\end {equation}
\begin {equation}
   \Delta {\cal I}_\pm =
   \frac{\zeta_\pm(2)-\zeta_\pm(3)}{2\pi}
   \left[ 
     \ln\left(\frac{T}{\md}\right)
     + \tfrac12 - \gammaE + \ln2
   \right]
   - \frac{\sigma_\pm}{2\pi} ,
\end {equation}
and
\begin {equation}
  \sigma_\pm \equiv
  \sum_{k=1}^\infty \frac{(\pm)^{k-1}}{k^3} \, \ln\bigl[(k{-}1)!\bigr] .
\label {eq:Spm}
\end {equation}
\end {subequations}
Above, $\gammaE$ is the Euler-Mascheroni constant.
As we'll discuss later, the sums
$\sigma_\pm$ are related to a certain generalization of the $\zeta$ function,
but we are unaware of any way of writing them in terms of more
usual mathematical constants.  Their
numerical values are
\begin {equation}
   \sigma_+ = 0.386043817389949\cdots
%       S_+ = 0.772087634779898\cdots
   \qquad\mbox{and}\qquad
   \sigma_- = 0.011216764589789\cdots \,.
%       S_- = 0.022433529179578\cdots \,.
\end {equation}

The result just quoted ignored running of the coupling constant.
If $\Lambda$ is so large that $g^2(\Lambda)$ is significantly
different from $g^2(\md)$ and $g^2(T)$, then one should make
the replacements
\begin {align}
   g^4 \ln\left(\frac{\Lambda}{\md}\right)
   &\to
   g^2(\Lambda) \, g^2(\md) \, \ln\left(\frac{\Lambda}{\md}\right) ,
\label {eq:runUVlog}
\\
   g^4 \ln\left(\frac{T}{\md}\right)
   &\to
   g^2(T) \, g^2(\md) \, \ln\left(\frac{T}{\md}\right) ,
\end {align}
following the discussion in Refs.\ \cite{ArnoldDogan,timelpm1,peshier},
and replace the $g^4$ which do not multiply logs by $g^4(T)$
as these constants turn out to be determined by the $q_\perp{\sim}T$
range of the $q_\perp$ integration in (\ref{eq:qhatdef}).
This compact prescription accounts
for 1-loop running of the coupling constant, provided
there are no vacuum mass thresholds between $\md$ and $\Lambda$.
The difference between $g^2(T)$ and $g^2(\md)$ is of order
$g^4 \ln (T/\md) \sim g^4 \ln (g^{-1})$ and so not significant
in the weak coupling limit, but no harm is done in accounting for
this particular higher-order correction to $\qhat$.

The replacement (\ref{eq:runUVlog}) has a finite limit as
$\Lambda \to \infty$:
\begin {equation}
   g^4 \ln\left(\frac{\Lambda}{\md}\right)
   \to
   \frac{g^2(\md)}{-2\bar\beta_0}
   \qquad
   (\Lambda = \infty) ,
\end {equation}
where
\begin {equation}
  \bar\beta_0 =
  - \frac{(11 \ca - 4 \Nf \tf)}{48 \pi^2}
  = - \frac{(33 - 2 \Nf)}{48 \pi^2}
\end {equation}
is the one-loop coefficient of the $\beta$ function for
$g^2(\mu) = [-\bar\beta_0 \ln(\mu^2/\Lambda^2)]^{-1}$.
So, if the effect of running is included, $\qhat(\infty)$ is finite
and given by the above formulas.
For applications to bremsstrahlung, however, we will generally want
$\qhat(\Lambda)$ for finite $\Lambda$.

% ---------------------------------------------------------------------------

\subsection {Result for gluon bremsstrahlung}

Consider bremsstrahlung of gluons with energy $xE \gg \md$ from a
high-energy particle of energy $E$ and species $s$.
The leading-log result for
bremsstrahlung from this particle
is (ignoring final state factors for the high-energy
particle and gluon)%
\footnote{
  The NLL analysis will be based on Arnold and Dogan
  \cite{ArnoldDogan}.  The leading-log result in (\ref{eq:brem})
  and (\ref{eq:LLmu}) of the current paper roughly corresponds to
  eqs.\ (1.1), (3.1a--b), and (4.16) of Ref.\ \cite{ArnoldDogan},
  with the correspondence that $\mu_\perp = \md \hat\mu_\perp$.
  The only difference is that we have written the answer in terms
  of the general $\qhat(Q_{\perp 0})$, whereas Ref. \cite{ArnoldDogan}
  specialized to the $q_\perp \ll T$ limit, resulting in their
  Eq.\ (4.12).  See Appendix \ref{app:NLL} of the current paper for
  more detail, including an explanation of differences in notation.
}
\begin {equation}
  \frac{d\Gamma_{s\to{\rm g}s}}{dx} =
  \frac{ \alpha \mu_\perp^2 \, P_{s\to{\rm g}}(x) }
       { 4 \pi \sqrt2 \, x(1-x) E} \,,
\label {eq:brem}
\end {equation}
where $x$ is the bremsstrahlung gluon momentum fraction,
$P_{s\to{\rm g}s}(x)$ is the usual vacuum splitting function,
and, at leading-log order,
\begin {equation}
  \mu_\perp^2 \simeq
  \Bigl\{
    8 x(1-x) E
    \left[ \half \ca + (C_s - \half \ca)x^2 + \half\ca(1-x)^2 \right]
    \qhat(Q_{\perp0})
  \Bigr\}^{1/2} .
\label {eq:LLmu}
\end {equation}
Here, $Q_{\perp0}$ is any rough guess (\ref{eq:Qperp})
of the total momentum transfer
during the formation time, and ambiguities in that guess of $O(1)$
factors only affect the answer beyond leading-log order.
The
Dokshitzer-Gribov-Lipatov-Altarelli-Parisi (DGLAP)
splitting functions in (\ref{eq:brem}) are
\begin {align}
   P_{{\rm q}\to {\rm g}}(x)
   &= \cf \, \frac{[1+(1-x)^2]}{x} \,,
\\
   P_{{\rm g}\to {\rm g}}(x)
   &= \ca \, \frac{[1 + x^4 + (1-x)^4]}{x(1-x)} \,.
\end {align}

If we now follow the same steps as Arnold and Dogan
\cite{ArnoldDogan}, the NLL result corresponds to
replacing (\ref{eq:LLmu}) by
\begin {multline}
  \mu_\perp^2 \simeq
  \bigl[
    8 x(1-x) E
  \bigr]^{1/2}
\\ \times
  \left\{
    \left[
      \half\ca \, \qhat( \xi^{1/2}\mu_\perp)
      +(C_s - \half\ca)x^2 \,
          \qhat\Bigl( \frac{\xi^{1/2}\mu_\perp}{x} \Bigr)
      +\half\ca (1-x)^2 \,
          \qhat\Bigl( \frac{\xi^{1/2}\mu_\perp}{1-x} \Bigr)
    \right]
  \right\}^{1/2} \!\!,
\label {eq:mu}
\end {multline}
where
\begin {equation}
   \xi \equiv \exp(2 - \gammaE + \tfrac{\pi}{4})
\label {eq:xi}
\end {equation}
is the same constant as in Arnold and Dogan,
and where (\ref{eq:mu}) is an implicit equation%
\footnote{
  As discussed in Sec.\ III of Arnold and Dogan \cite{ArnoldDogan},
  one can choose to solve this equation iteratively, starting from
  some initial guess $Q_{\perp 0}$ and explicitly generating the first two
  terms of the expansion in inverse logs.  However, as discussed
  in Ref.\ \cite{ArnoldDogan}, the implicit form has the advantage
  that one does not have to generate an initial guess $Q_{\perp 0}$.
}
for $\mu_\perp$.
Details are given in Appendix \ref{app:NLL}.

The assumption that goes into this result is that
$\qhat(\Lambda)$ is proportional to $\ln\Lambda$ at the scale of
its arguments---that is, that
$d\Gamma_{\rm el}/d^2 q_\perp \propto 1/q_\perp^4$ for $q_\perp$ of
order the argument $\Lambda$ of $\qhat$.
This assumption works for arguments in the range
$\md \ll \Lambda \ll T$ as well as the case $T \ll \Lambda$
considered in this paper.
As a result, if one uses
the small $q_\perp$ form (\ref{eq:qhatsmall}) of
$\qhat(\Lambda)$, then the formula
(\ref{eq:mu}) reproduces the result found in Arnold and Dogan.
For high energy particles with parametrically
$E \gg T/g^4 \ln(1/g)$, however, one should instead
use (\ref{eq:qhateqs}) for $\qhat$.

As discussed in Refs.\ \cite{ArnoldDogan,timelpm1,BDMPS3},
if the running of the coupling is included, the explicit factor
of $\alphas$ in (\ref{eq:brem}) should plausibly
be $\alphas(\mu_\perp)$.

Our result (\ref{eq:qhateqs}) formally treated $\md \ll T$.
If one evaluated $\qhat(\Lambda)$ in a way that relaxed this
assumption, Eqs.\ (\ref{eq:brem}) and
(\ref{eq:mu}) could still be used to calculate
the bremsstrahlung rate for high-energy particles.

% ----------------------------------------------------------------------------

\subsection {Result for Pair Production}

Pair production is the same as gluon bremsstrahlung except for (i)
a change in group factors, and (ii) use of the corresponding
vacuum splitting function
\begin {equation}
   P_{{\rm g}\to {\rm q}}(x)
   = \Nf \tf [x^2+(1-x)^2] .
\end {equation}
See, for example, the more symmetric presentation given in
Arnold and Dogan \cite{ArnoldDogan}.
The result (summed over quark flavors) is
\begin {equation}
  \frac{d\Gamma_{{\rm g}\to{\rm q}\bar{\rm q}}}{dx} =
  \frac{ \alpha \mu_\perp^2 \, P_{{\rm g}\to{\rm q}}(x) }
       { 4 \pi \sqrt2 \, x(1-x) E}
\end {equation}
with
\begin {multline}
  \mu_\perp^2 \simeq
  \bigl[
    8 x(1-x) E
  \bigr]^{1/2}
\\ \times
  \left\{
    \left[
      (\cf - \half \ca) \, \qhat( \xi^{1/2}\mu_\perp)
      +\half\ca x^2 \,
          \qhat\Bigl( \frac{\xi^{1/2}\mu_\perp}{x} \Bigr)
      +\half\ca (1-x)^2 \,
          \qhat\Bigl( \frac{\xi^{1/2}\mu_\perp}{1-x} \Bigr)
    \right]
  \right\}^{1/2} \!\!,
\end {multline}

%============================================================================

\section {\boldmath$d\Gamma_{\rm el}/d^2q_\perp$ and \boldmath$\qhat(\Lambda)$}
\label {sec:qhat}

\subsection {Strategy}

Our first goal is to determine how the differential elastic scattering
rate $d\Gamma_{\rm el}/d^2q_\perp$ interpolates at $q_\perp \sim T$
between the two limits shown in (\ref{eq:dsig}).
For $q_\perp \gg \md$, screening effects can be ignored, and the
differential cross-section will have the form
\begin {equation}
   \frac{d\Gamma_{\rm el}}{d^2q_\perp}
   \simeq
   \frac{C_R}{(2\pi)^2} \times
      \displaystyle\frac{g^4 T^3 \, F(q_\perp/T)}{q_\perp^4}
\end {equation}
for some function $F(q_\perp/T)$ with
\begin {equation}
   g^4 T^3 \, F(0) = g^2 T \md^2
   \qquad \mbox{and} \qquad
   g^4 T^3 \, F(\infty) = g^4 {\cal N} .
\label {eq:Flimits}
\end {equation}
In the limit of weak coupling, once we know $F(q_\perp/T)$
we can then construct a formula valid
to leading-order at all scales for $q_\perp$ as
\begin {equation}
   \frac{d\Gamma_{\rm el}}{d^2q_\perp}
   \simeq
   \frac{C_R}{(2\pi)^2} \times
      \frac{g^4 T^3 \, F(q_\perp/T)}{q_\perp^2 (q_\perp^2+\md^2)}
\label {eq:dsigF}
\end {equation}
since $\md \ll T$.

It will be convenient to write $g^4 T^3 F$ in the form
\begin {equation}
   g^4 T^3 \, F(q_\perp/T)
   = \bigl[\num_{\rm b} \, I_+(q_\perp/T)
          +\num_{\rm f} \, I_-(q_\perp/T)\bigr]
     \frac{g^4 T^3}{\pi^2} \,.
\label {eq:FIpm}
\end {equation}
The functions $I_\pm(q_\perp/T)$, to be determined,
extrapolate between
\begin {equation}
   I_\pm(0) = \zeta_\pm(2)
   \qquad \mbox{and} \qquad
   I_\pm(\infty) = \zeta_\pm(3) .
\label {eq:Ipmlimits}
\end {equation}
The equivalence to (\ref{eq:Flimits}) can be seen
from Eqs.\ (\ref{eq:mdN}) for $\md$ and ${\cal N}$.

% ---------------------------------------------------------------------------

\subsection {Starting Point for \boldmath$d\Gamma_{\rm el}/d^2q_\perp$}

In general, the rate for a high-energy particle of energy $E$
to scatter from the plasma is given by
\begin {equation}
   \frac{d\Gamma_{{\rm el},s}}{d^2q_\perp}
   \simeq
   \int \frac{dq_z}{(2\pi)^3}
   \sum_{s_2} d_{R_2} \bar\nu_{s_2}
   \int \frac{d^3 p_2}{(2\pi)^3} \>
   \frac{d\sigma_{\rm el}}{d^3q}
   \,
   f_{s_2}(\p_2) \, \bigl[ 1 \pm f_{s_2}(\p_2{-}\q) \bigr] ,
\label {eq:dsig0}
\end {equation}
where $z$ is the direction of motion of the high-energy particles,
$\p_2$ is the momentum of a particle in the plasma of species
$s_2$, $\sigma_{\rm el}$ is the cross-section for scattering from that plasma
particle, $f(\p_2)$ is a Bose or Fermi distribution that
accounts for the probability of encountering the plasma particle,
and $1 \pm f(\p_2{-}\q)$ is a final-state Bose enhancement or Fermi
blocking factor for the plasma particle after transferring momentum
$\q$ to the high-energy particle.  We assume that $E \gg T$ and so
do not need to include any final state factor for the high energy
particle.
In (\ref{eq:dsig0}), $\bar\nu_{s_2}$ is the number of spin and flavor
degrees of freedom for species $s_2$ (2 for gluons, $4\Nf$ for the sum
of quarks and anti-quarks).

For high-energy particles (in this case, $E \gg \md$), elastic
scattering from the plasma is dominated by $t$-channel gluon exchange.
The infrared behavior of $t$-channel gluon exchange is cut off in
the infrared at $q_\perp \sim \md$
by the effects of Debye screening and related phenomena
in the plasma.
To leading order
in the weak-coupling limit, the problem is simplified by the fact that
we can ignore screening effects when investigating $q_\perp \sim T$
since $T \gg \md$.
To leading order, the differential elastic scattering rate for
$q_\perp \gg \md$ is then
\begin {multline}
   \frac{d\Gamma_{{\rm el}}}{d^2q_\perp}
   \simeq
   \int \frac{dq_z \> d\omega}{(2\pi)^4} \>
   \sum_{s_2} \bar\nu_{s_2}
   \int
   \frac{d^3 p_2}{(2\pi)^3} \>
     \frac{C_R t_{R_2} g^4}{(2p)^2 \, 2p_2 \, 2|\p_2-\q|}
     \left| \frac{4 P_\mu P_2^\mu}{Q^2} \right|^2
     f_{s_2}(\p_2) \, \bigl[ 1 \pm f_{s_2}(\p_2-\q) \bigr]
\\ \times
   2\pi \, \delta(\omega - q_z)
   \,
   2\pi \, \delta(\omega + |\p_2-\q| - p_2)
\label {eq:dsig2}
\end {multline}
for a massless high-energy particle with momentum $\p$
in the $z$ direction.
We use capital letters for 4-momenta, with $P=(p,\p)$ and
$Q \equiv (\omega,\q)$.
The factors of $(2 p)^{-1}$, $(2p)^{-1}$, $(2p_2)^{-1}$, and
$(2|\p_2-\q|)^{-1}$ are the usual initial and final state relativistic
phase space normalizations, where we've taken the high energy limit
$E \gg q$.
The $\delta(\omega + |\p_2-\q| - p_2)$ is energy conservation for
the (massless) plasma particle.  The other $\delta$-function is
energy conservation $\delta(|\p+\q| - \omega - p)$ for the
incident high-energy particle, again taking the high-energy limit
$p \gg q$.  The two gluon vertices in the amplitude are
$2 g P^\mu$ and $2 g P_2^\mu$ (times color generators),
regardless of the types of particles
colliding.  The appearance of this universal
form can be understood as a consequence that, in the high-energy
limit $E \gg T$, the upper limit
(\ref{eq:Qperp}) for the range of individual
$q_\perp$ transfers dominating
bremsstrahlung is small compared to the center-of-mass energy
$\sim (ET)^{1/2}$ for a collision with a plasma particle.
In the center-of-mass frame, the $t$-channel gluon is soft
compared to either particle involved in the elastic collision,
and so we may use the universal form that gluon-particle vertices
take in the soft gluon limit.  Note that $P_\mu P_2^\mu$ in
(\ref{eq:dsig2}) could have equally well been written
using the final plasma particle momentum $P_2-Q$ as
$P_\mu (P_2-Q)^\mu$, due to the $\delta(\omega-q_z)$.

Performing the $\omega$ integration in (\ref{eq:dsig2}),
\begin {multline}
   \frac{d\Gamma_{{\rm el}}}{d^2q_\perp}
   \simeq
   \frac{C_R}{(2\pi)^2 q_\perp^4}
   \sum_{s_2} \bar\nu_{s_2}
   t_{R_2} g^4
   \int \frac{dq_z}{2\pi} \>
   \int\frac{d^3 p_2}{(2\pi)^3} \>
     \frac{(p_2-p_{2z})^2}{p_2 |\p_2-\q|} \,
     f_{s_2}(\p_2) \, \bigl[ 1 \pm f_{s_2}(\p_2-\q) \bigr]
\\ \times
   2\pi \, \delta(q_z + |\p_2-\q| - p_2) .
\label {eq:dsig3}
\end {multline}
In thermal equilibrium at zero chemical potential,
the distributions $f_{s_2}$ are all the same
for massless bosons, and also all the same for massless fermions,
and we can rewrite
$d\Gamma_{\rm el}/d^2q_\perp$ as
\begin {equation}
   \frac{d\Gamma_{{\rm el}}}{d^2q_\perp}
   \simeq
   \frac{C_R}{(2\pi)^2 q_\perp^4}
   \times
   \bigl[\num_{\rm b} \, I_+(q_\perp/T)
        +\num_{\rm f} \, I_-(q_\perp/T)\bigr]
   \frac{g^4 T^3}{\pi^2}
\end {equation}
with
\begin {multline}
   I_\pm(q_\perp/T) = \frac{\pi^2}{T^3}
   \int \frac{dq_z}{2\pi} \>
   \int\frac{d^3 p_2}{(2\pi)^3} \>
     \frac{(p_2-p_{2z})^2}{p_2 |\p_2-\q|} \,
     f_\pm(\p_2) \, \bigl[ 1 \pm f_\pm(\p_2-\q) \bigr]
\\ \times
   2\pi \, \delta(q_z + |\p_2-\q| - p_2) .
\label {eq:Ipmintegral}
\end {multline}

% ---------------------------------------------------------------------------

\subsection {Recasting \boldmath$d\Gamma_{\rm el}/d^2q_\perp$ as a
             double sum}
\label {sec:doublesum}

Now rewrite the equilibrium Bose and Fermi distribution functions as
sums of exponentials,%
\footnote{
  The analysis of Ref.\ \cite{ArnoldDogan} could also be applied to
  non-equilibrium isotropic situations.  Here, however, we are
  specializing to equilibrium distributions.
}
\begin {equation}
   f_\pm(p) = \frac{1}{e^{\beta p}\mp 1}
   = \sum_{m=1}^\infty (\pm)^{m-1} e^{-m\beta p} ,
\end {equation}
\begin {equation}
   1 \pm f_\pm(p)
   = \sum_{n=0}^\infty (\pm)^n e^{-n\beta p} ,
\end {equation}
with $\beta \equiv 1/T$.
Then
\begin {equation}
   I_\pm(q_\perp/T) =
   \sum_{m=1}^\infty \sum_{n=0}^\infty (\pm)^{m+n-1}
   I_{mn}(q_\perp/T)
\label {eq:Ipm}
\end {equation}
with
\begin {equation}
   I_{mn}(q_\perp/T) \equiv
   \frac{\pi^2}{T^3}
   \int \frac{dq_z}{2\pi}
   \frac{d^3 p_2}{(2\pi)^3} \>
   \frac{(p_2-p_{2z})^2}{p_2 |\p_2-\q|} \,
   e^{-m\beta p_2} e^{-n\beta|\p_2-\q|} \,
   2\pi \, \delta(q_z + |\p_2-\q| - p_2) .
\label {eq:Imn0}
\end {equation}
This integral is evaluated in Appendix \ref{app:Imn} and yields
\begin {equation}
   I_{mn}(q_\perp/T) =
   \frac{mn}{2(m+n)^3} \,
   \left(\frac{q_\perp}{T}\right)^2
   K_2\Bigl( \frac{q_\perp}{T} \sqrt{mn} \Bigr) ,
\label {eq:Imn}
\end {equation}
where $K_\nu(z)$ is the modified Bessel function of the second kind.
The case $n=0$ gives the $n\to0$ limit of the above formula,
\begin {equation}
   I_{m0} =
   \frac{1}{m^3} \,.
\end {equation}

In the $q_\perp \to 0$ limit, $I_{mn} \to 1/(m+n)^3$, and so
\begin {equation}
   I_\pm(0) =
   \sum_{m=1}^\infty \sum_{n=0}^\infty
   \frac{(\pm)^{m+n-1}}{(m+n)^3}
   =
   \sum_{m+n=1}^\infty 
   \frac{(\pm)^{m+n-1}}{(m+n)^2}
   =
   \zeta_\pm(2) .
\label {eq:Ipm0}
\end {equation}
In the opposite limit of $q_\perp \to \infty$, only the $n=0$ terms
survive in the double sum (\ref{eq:Ipm}), giving
\begin {equation}
   I_\pm(\infty) =
   \sum_{m=1}^\infty (\pm)^{m-1} I_{m0}
   = \zeta_{\pm}(3) .
\label {eq:Ipminfty}
\end {equation}
These two limits are in accord with (\ref{eq:Ipmlimits}).
We will later find it useful to extract the $n=0$ contribution from
the general case, writing
\begin {subequations}
\label {eq:Ipmformula}
\begin {equation}
   I_\pm(q_\perp/T) = I_\pm(\infty) + \Delta I_\pm(q_\perp/T)
   = \zeta_\pm(3) + \Delta I_\pm(q_\perp/T)
\end {equation}
with
\begin {equation}
   \Delta I_\pm(q_\perp/T) =
   \sum_{m,n=1}^\infty (\pm)^{m+n-1}
   I_{mn}(q_\perp/T) .
\label {eq:DeltaIpm}
\end {equation}
\end {subequations}

If one wished to evaluate the functions $I_\pm(q_\perp/T)$ numerically,
the expansion (\ref{eq:DeltaIpm}) converges rapidly for
$q_\perp \gtrsim T$.  For the case of small $q_\perp$, we show in
Appendix \ref{app:smallqt} that the expansion is%
\footnote{
  A numerically modest
  formula which reproduces $I_+({\cal Q})$ to a few tenths of a
  percent is to (i) for ${\cal Q} < 3.2$ use (\ref{eq:Ipsmall})
  plus $-0.0062 \, {\cal Q}^3$ inside the square brackets, and
  (ii) for ${\cal Q} > 3.2$ use the $m+n \le 3$
  terms of (\ref{eq:Ipmformula}).  For $I_-({\cal Q})$ do the same, but
  with $-0.0031 \, {\cal Q}^3$ added inside the square brackets of
  (\ref{eq:Imsmall}).
}
\begin {align}
   I_+({\cal Q}) &=
   \zeta_+(2) \left[
     1 - \tfrac3{16} {\cal Q}
     + ( \tfrac1{24} + \tfrac{1}{8\pi^2} ) {\cal Q}^2
     + O({\cal Q}^3)
   \right] ,
\label {eq:Ipsmall}
   \\
   I_-({\cal Q}) &=
   \zeta_-(2) \left[
     1
     + ( \tfrac1{24} - \tfrac{1}{4\pi^2} ) {\cal Q}^2
     + O({\cal Q}^3)
   \right] .
\label {eq:Imsmall}
\end {align}

% -------------------------------------------------------------------------

\subsection{Integrating to get \boldmath$\hat q$}

Using (\ref{eq:dsigF}) and (\ref{eq:FIpm}),
our leading-order formula for the differential elastic scattering rate
is
\begin {equation}
   \frac{d\Gamma_{\rm el}}{d^2q_\perp}
   \simeq
   \frac{C_R}{(2\pi)^2}
      \frac{g^4 T^3
            \bigl[\num_{\rm b} \, I_+(q_\perp/T)
                 +\num_{\rm f} \, I_-(q_\perp/T)
            \bigr]}
           {\pi^2 q_\perp^2 (q_\perp^2+\md^2)}
   \,.
\end {equation}
So, to compute the integral (\ref{eq:qhatdef}) that gives
$\qhat(\Lambda)$, we turn to evaluating the integrals
\begin {equation}
   {\cal I}_\pm(\Lambda) \equiv
   \int_{q_\perp<\Lambda} \frac{d^2 q_\perp}{(2\pi)^2} \>
   \frac{I_\pm(q_\perp/T)}{(q_\perp^2+\md^2)} \,.
\end {equation}
In the weak coupling limit, one can choose a momentum scale $\lambda$
between $\md$ and $T$, with $\md \ll \lambda \ll T$.  Then
one may split the integral at
$\lambda$ into separate pieces for which $q_\perp \ll T$ or
$q_\perp \gg \md$ approximations may be made:
\begin {equation}
   {\cal I}_\pm(\Lambda) \simeq
   \int_{q_\perp<\lambda} \frac{d^2 q_\perp}{(2\pi)^2} \>
   \frac{I_\pm(0)}{(q_\perp^2+\md^2)}
   +
   \int_{\lambda<q_\perp<\Lambda} \frac{d^2 q_\perp}{(2\pi)^2} \>
   \frac{I_\pm(q_\perp/T)}{q_\perp^2} \,.
\label {eq:qhatdome}
\end {equation}
In Appendix \ref{app:qhat}, we then evaluate these integrals
using the double sum (\ref{eq:Ipmformula}) for the last one.
To leading order, we obtain the result (\ref{eq:qhateqs})
with
\begin {equation}
  \sigma_\pm =
  \half \sum_{m,n=1}^\infty (\pm)^{m+n-1}
  \frac{\ln(mn)}{(m+n)^3}
  =
  \sum_{m,n=1}^\infty (\pm)^{m+n-1}
  \frac{\ln(m)}{(m+n)^3} \,.
\label {eq:Spm2}
\end {equation}
This can be written as 
$\sigma_+ = - \partial_a T(a,0,3) \bigl|_{a=0}$ where
\begin {equation}
   T(a,b,c) \equiv \sum_{m,n=1}^\infty \frac{1}{m^a n^b (m+n)^c}
\end {equation}
is the
Tornheim zeta function \cite{tornheim},
and similarly $\sigma_-$ can be defined
in terms of its generalization.  However, this does not seem to give
any information that is more useful than the sum itself.
By letting $k=m+n$ and then summing over $n$ for fixed $k$, one
obtains the single-sum formula (\ref{eq:Spm}) for $\sigma_\pm$.

% =========================================================================

\begin{acknowledgments}

We are indebted to Simon Caron-Huot and Guy Moore for useful conversations.
It was Simon Caron-Huot's work \cite{simon} in particular that made us
understand that our NLL bremsstrahlung rate should be presented in terms
of a UV-regulated $\hat q$.
This work was supported, in part, by the U.S. Department
of Energy under Grant No.~DE-FG02-97ER41027.

\end{acknowledgments}

% =========================================================================

\appendix

\section{\boldmath$I_{mn}$}
\label {app:Imn}

In the integral (\ref{eq:Imn0}) defining $I_{mn}$, use the
$\delta$-function to rewrite
\begin {equation}
   (p_2-p_{2z})^2 =
   (p_2-p_{2z}) \bigl[ |\p_2-\q| - (p_{2z}-q_z) \bigr] .
\end {equation}
Then rewrite the $\delta$-function as
\begin{equation}
  2\pi \, \delta(q_z + |\p_2-\q| - p_2)
  = \int_{-\infty}^{+\infty} d\lambda \> e^{i\lambda(q_z + |\p_2-\q|-p_2)} .
\end {equation}
The $\p_2$ integral now takes the form of a convolution of
functions $W_m$ and $W_n^*$ defined by
\begin {equation}
  W_m(\p_2,\lambda) \equiv
  \frac{(p_2-p_{2z})}{p_2} \, e^{-(m \beta+i\lambda) p_2} .
\label {eq:Wdef}
\end {equation}
To turn the convolution into a simple product, we Fourier transform
from $\p_2$ to $\r$:
\begin {equation}
   I_{mn}(q_\perp/T) \equiv
   \frac{\pi^2}{T^3}
   \int \frac{dq_z}{2\pi}
   \int d\lambda \> e^{i \lambda q_z}
   \int d^3r \>
   \tilde W_m(\r,\lambda) \, \tilde W_n^*(\r,\lambda) \, e^{-i\q\cdot\r} .
\end {equation}
The $q_z$ integral then gives
$\delta(\lambda-z)$, which can be used to do the
$\lambda$ integral:
\begin {equation}
   I_{mn}(q_\perp/T) \equiv
   \frac{\pi^2}{T^3}
   \int d^3r \>
   \tilde W_m(\r,z) \, \tilde W_n^*(\r,z) \, e^{-i\q_\perp\cdot\r_\perp} .
\label {eq:Imn2}
\end {equation}
The Fourier transform of (\ref{eq:Wdef}) evaluated at $\lambda=z$
is
\begin {equation}
  \tilde W_m(\r,z) =
  \frac{T^3}{\pi^2} \, \frac{m}{[m^2 + 2imzT + (r_\perp T)^2]^2} \,.
\end {equation}
Now do the $z$ integration in (\ref{eq:Imn2})
by closing the contour in the upper-half complex plane
and picking up the double pole there:
\begin {equation}
   I_{mn}(q_\perp/T) \equiv
   \frac{2 T^2}{\pi}
   \int d^2r_\perp \>
   \frac{(mn)^2}{(m+n)^3[(r_\perp T)^2 + mn]^3}
   \, e^{-i\q_\perp\cdot\r_\perp} .
\end {equation}
Finally, performing the $\r_\perp$ integral gives
(\ref{eq:Imn}).

% =========================================================================

\section{Small \boldmath$q_\perp$ expansion of \boldmath$I_\pm$}
\label{app:smallqt}

It is possible to find the small {$\cal Q$} expansion of
$I_\pm({\cal Q})$ starting directly from the integral formula
(\ref{eq:Ipmintegral}).  For the bosonic case, at least, we find
it easier to instead start from the double sum formula derived in
Sec.\ \ref{sec:doublesum}.

\subsection {The \boldmath$O(q_\perp)$ piece of \boldmath$I_+$}

Start from the double sum of (\ref{eq:Ipm}) and (\ref{eq:Imn})
and subtract off the $q_\perp = 0$ piece:
\begin {equation}
   \delta I_+ \equiv I_+({\cal Q}) - I_+(0)
   = 
   \sum_{m,n=1}^\infty
   \frac{
     \left[
       \half {\cal Q}^2 m n \, K_2( {\cal Q} \sqrt{mn} )
       - 1
     \right]
   }{(m+n)^3}
   \,.
\label {eq:Ipsmall2}
\end {equation}
For small ${\cal Q} \equiv q_\perp/T$, this sum is dominated by
large $m$ and $n$, and so we can replaced the sum by an integral:
\begin {equation}
   \delta I_+
   \simeq
   \int_0^\infty dm \> dn \>
   \frac{
     \left[
       \half {\cal Q}^2 m n \, K_2( {\cal Q} \sqrt{mn} )
       - 1
     \right]
   }{(m+n)^3}
   \,.
\end {equation}
Change integration variable from $n$ to $x \equiv {\cal Q} \sqrt{m n}$,
and then do the $m$ integral to get
\begin {equation}
   \delta I_+
   \simeq
   \tfrac{\pi}{16} \, {\cal Q}
   \int_0^\infty dx \>
   \left[
     K_2(x) - \frac{2}{x^2}
   \right]
   = -\tfrac{\pi^2}{32} \, {\cal Q}.
\end {equation}
This gives the $O({\cal Q})$ term in (\ref{eq:Ipsmall}).

% --------------------------------------------------------------------------

\subsection {The \boldmath$O(q_\perp^2)$ piece of \boldmath$I_+$}

The difference between the double sum and the integral approximation
made above is
\begin {equation}
   \delta^2 I_+
   = 
   \biggl( \sum_{m,n=1}^\infty - \int_0^\infty dm\>dn \biggr)
   \frac{
     \left[
       \half {\cal Q}^2 m n \, K_2( {\cal Q} \sqrt{mn} )
       - 1
     \right]
   }{(m+n)^3}
   \, .
\end {equation}
Now rewrite
\begin {equation}
%   \sum_{m,n} - \int_0^\infty dm\>dn
   \sum_m \sum_n - \int_m \int_n
   =
   \biggl( \sum_m - \int_m \biggr) \int_n
   + \int_m \biggl( \sum_n - \int_n \biggr)
   + \biggl( \sum_m - \int_m \biggr) \biggl( \sum_n - \int_n \biggr) ,
\end {equation}
and correspondingly (using the $m \leftrightarrow n$ symmetry of
the sum for $I_+$)
\begin {equation}
   \delta^2 I_+ =
   \delta^{2{\rm a}} I_+
   + \delta^{2{\rm a}} I_+
   + \delta^{2{\rm b}} I_+ .
\label {eq:Ipexpand}
\end {equation}

To evaluate $\delta^{2{\rm a}} I_+$, again change integration variable
from $n$ to $x \equiv {\cal Q}\sqrt{mn}$ to get
\begin {equation}
  \delta^{2{\rm a}} I_+ =
  {\cal Q}^4
  \biggl( \sum_{m=1}^\infty - \int_0^\infty dm \biggr)
  F(m;{\cal Q})
\end {equation}
where
\begin {equation}
  F(m;{\cal Q}) \equiv
  2 m^2
  \int_0^\infty
  \frac{dx \> x [ \half x^2 \, K_2(x) - 1 ]}{(x^2+m^2{\cal Q}^2)^3}
  .
\label {eq:Fm}
\end {equation}
For small ${\cal Q}$, $F(m;{\cal Q})$ is a slowly varying function of $m$,
which in general means that
\begin {equation}
   \biggl( \sum_{m=1}^\infty - \int_0^\infty dm \biggr) F(m;{\cal Q})
   \simeq - \half F(0;{\cal Q}) .
\end {equation}
In our case, $F(0;{\cal Q})$ should be understood to be the
$m \to 0$ limit $-1/8{\cal Q}^2$ of (\ref{eq:Fm}), which then gives
\begin {equation}
  \delta^{2{\rm a}} I_+ =
  \tfrac1{16} {\cal Q}^2 .
\label {eq:Ip2a}
\end {equation}

The other term $\delta^{2{\rm b}}I_+$ in (\ref{eq:Ipexpand})
will be turn out to be dominated by $m$ and $n$ with
$mn \ll 1/{\cal Q}^2$ in the small ${\cal Q}$ limit.
Making this small ${\cal Q}$ approximation to the argument of
$K_2$, we get
\begin {equation}
  \delta^{2{\rm b}} I_+ =
  - {\cal Q}^2
  \biggl( \sum_{m=1}^\infty - \int_0^\infty dm \biggr)
  \biggl( \sum_{n=1}^\infty - \int_0^\infty dn \biggr)
  G(m,n)
\end {equation}
with
\begin {equation}
  G(m,n) = \frac{m n}{4(m+n)^3} .
\label {eq:Gmn}
\end {equation}
Now rewrite
\begin {multline}
  \biggl( \sum_{m=1}^\infty - \int_0^\infty dm \biggr)
  \biggl( \sum_{n=1}^\infty - \int_0^\infty dn \biggr)
  G(m,n)
  =
\\
  \sum_{m,n=1}^\infty
  \biggl\{
    G(m,n)
    - \int_{m-1}^m dm' \> G(m',n)
    - \int_{n-1}^n dn' \> G(m,n')
    + \int_{m-1}^m dm' \int_{n-1}^n dn' \> G(m',n')
  \biggr\} .
\end {multline}
Explicit evaluation of the integrals using (\ref{eq:Gmn}) yields
\begin {equation}
  \delta^{2{\rm b}} I_+ =
  {\cal Q}^2
  \sum_{m,n=1}^\infty
  \frac{(m^3-3m^2 n - 3m n^2 + n^3)+4mn}{8(m+n)^3(m+n-1)^2(m+n-2)} \,,
\label {eq:d2b}
\end {equation}
where the summand for $m=n=1$ should be treated as the limiting
value $-\tfrac1{32}$.
Defining $k \equiv m+n$, (\ref{eq:d2b}) can be rewritten
\begin {equation}
  \delta^{2{\rm b}} I_+ =
  {\cal Q}^2
  \sum_{k=2}^\infty \sum_{n=1}^{k-1}
  \frac{k^3-6k^2 n + 6kn^2 + 4 kn - 4n^2}{8k^3(k-1)^2(k-2)}
 \,.
\end {equation}
Doing the $n$ sum first (and treating the $k=2$ case separately,
which does not fit the pattern of $k>2$):
\begin {equation}
  \delta^{2{\rm b}} I_+ =
  - {\cal Q}^2
  \biggl[
    \tfrac1{32} +
    \sum_{k=3}^\infty
    \frac{1}{24k^2(k-1)}
  \biggr]
  =
  - {\cal Q}^2 \bigl[
    \tfrac5{48} - \tfrac{1}{24} \, \zeta(2)
  \bigr] .
\label {eq:Ip2b}
\end {equation}
Putting (\ref{eq:Ip2a}) and (\ref{eq:Ip2b}) into
(\ref{eq:Ipexpand}) then yields
the $O({\cal Q}^2)$ term in (\ref{eq:Ipsmall}).

% --------------------------------------------------------------------------

\subsection {The \boldmath$O(q_\perp^2)$ piece of \boldmath$I_-$}

The case of $I_-$ is much simpler because there is no
$O({\cal Q})$ term in the expansion.  A quick but non-rigorous
way to obtain the answer is to naively expand the summand of
(\ref{eq:Ipm}) and (\ref{eq:Imn}) in
powers of ${\cal Q}$:
\begin {align}
  \delta^2 I_- &=
  - {\cal Q}^2 \sum_{m,n=1}^\infty
  (-)^{m+n-1} \frac{mn}{4(m+n)^3}
\nonumber\\
  &=
  - {\cal Q}^2 \sum_{k=2}^\infty \sum_{n=1}^{k-1}
  (-)^{k-1} \frac{(k-n)n}{4k^3}
\nonumber\\
  &=
  - \tfrac1{24} {\cal Q}^2 \sum_{k=2}^\infty
  (-)^{k-1} \left(1 - \frac{1}{k^2}\right) .
\end {align}
If one then interprets $\sum (-)^{k-1}$ as being $\zeta_-(0) = \half$, then
\begin {equation}
  \delta^2 I_- = - \tfrac{1}{24} [\zeta_-(0) - \zeta_-(2)] {\cal Q}^2 ,
\label {eq:Ipexpand2}
\end {equation}
where $\zeta_-(2) = \pi^2/12.$  This gives the result
(\ref{eq:Imsmall}) quoted previously.

A more reliable way to make the same calculation is to start with
the convergent, un-expanded sum analogous to (\ref{eq:Ipsmall2}),
\begin {equation}
   I_-({\cal Q}) - I_-(0)
   = 
   \sum_{m,n=1}^\infty
   (-)^{m+n-1} H(m,n;{\cal Q}) ,
\end {equation}
\begin {equation}
   H(m,n;{\cal Q}) \equiv
   \frac{
     \left[
       \half {\cal Q}^2 m n \, K_2( {\cal Q} \sqrt{mn} )
       - 1
     \right]
   }{(m+n)^3}
   \,.
\end {equation}
Now block the sum into $2\times2$ blocks as
\begin {multline}
   I_-({\cal Q}) - I_-(0)
   = 
   \sum_{m,n~{\rm odd}}
   \bigl[
      - H(m,n;{\cal Q}) + H(m{+}1,n;{\cal Q})
\\
      + H(m,n{+}1;{\cal Q}) - H(m{+}1,n{+}1;{\cal Q})
   \bigr]
   .
\end {multline}
At this point, one can safely expand the summand to
order ${\cal Q}^2$.  Then change
summation variables from $m$ to $k=m+n$, and then sum over first
$n$ and then $k$.  The final result is (\ref{eq:Ipexpand2}).

% =========================================================================

\section{\boldmath${\cal I}_\pm$}
\label{app:qhat}

In this appendix, we evaluate the integrals in (\ref{eq:qhatdome}).
The non-trivial integral is the second one:
\begin {equation}
   {\cal I}^{(2)} \equiv
   \int_{\lambda<q_\perp<\Lambda} \frac{d^2 q_\perp}{(2\pi)^2} \>
   \frac{I_\pm(q_\perp/T)}{q_\perp^2}
   =
   \frac{I_\pm(\infty)}{2\pi} \, \ln\left( \frac{\Lambda}{\lambda} \right) 
   + 
   \int_{\beta\lambda<{\cal Q}<\beta\Lambda} \frac{d^2{\cal Q}}{(2\pi)^2} \>
   \frac{\Delta I_\pm({\cal Q})}{{\cal Q}^2} \,,
\end {equation}
where the last equality uses (\ref{eq:Ipmformula}) and switches to the
dimensionless integration variable ${\cal Q} \equiv q_\perp/T$.
Because $\Delta I_\pm$ falls off for ${\cal q} \to \infty$, we can
drop the UV regularization ${\cal Q}<\beta\Lambda$ in the last
integral.

It's useful to now change infrared regularization by inserting
an initially unnecessary factor of ${\cal Q}^{2\epsilon}$
with the limit $\epsilon \to 0^+$ taken at the end of the day.
Then rewrite the above as
\begin {equation}
   {\cal I}_\pm^{(2)} =
   \frac{I_\pm(\infty)}{2\pi} \, \ln\left( \frac{\Lambda}{\lambda} \right) 
   + 
   \int \frac{d^2{\cal Q}}{(2\pi)^2} \>
   \frac{\Delta I_\pm({\cal Q})}{{\cal Q}^{2(1-\epsilon)}}
   -
   \int_{{\cal Q}<\beta\lambda} \frac{d^2{\cal Q}}{(2\pi)^2} \>
   \frac{\Delta I_\pm({\cal Q})}{{\cal Q}^{2(1-\epsilon)}} \,.
\label {eq:Ipm2}
\end {equation}
In the last term, we can replace $\Delta I_\pm({\cal Q})$ by
$\Delta I_\pm(0)$, giving
\begin {equation}
   \int_{{\cal Q}<\beta\lambda} \frac{d^2{\cal Q}}{(2\pi)^2} \>
   \frac{\Delta I_\pm({\cal Q})}{{\cal Q}^{2(1-\epsilon)}}
   \simeq
   \Delta I_\pm(0) \, \frac{(\beta\lambda)^{2\epsilon}}{4\pi\epsilon}
   =
   \frac{\Delta I_\pm(0)}{4\pi} \left[
      \frac{1}{\epsilon} + 2\ln\left(\frac{\lambda}{T}\right)
      + O(\epsilon)
   \right] .
\label {eq:int1}
\end {equation}
For the other integral in (\ref{eq:Ipm2}), we use the double sum
formula of (\ref{eq:DeltaIpm}) and (\ref{eq:Imn})
for $\Delta I_\pm$:
\begin {align}
   \int \frac{d^2{\cal Q}}{(2\pi)^2} \>
   \frac{\Delta I_\pm({\cal Q})}{{\cal Q}^{2(1-\epsilon)}}
   &=
   \sum_{m,n=1}^\infty (\pm)^{m+n-1}
   \frac{mn}{2(m+n)^3} \,
   \int \frac{d^2{\cal Q}}{(2\pi)^2} \>
   {\cal Q}^{2\epsilon}
   K_2\Bigl( {\cal Q} \sqrt{mn} \Bigr)
\nonumber\\
   &=
   \sum_{m,n=1}^\infty (\pm)^{m+n-1}
   \frac{(mn)^{-\epsilon}}{4\pi(m+n)^3} \,
   \int_0^\infty dx \>
   x^{1+2\epsilon}
   K_2(x) .
\end {align}
The last integral gives
$2^{2\epsilon} \, \Gamma(\epsilon) \, \Gamma(2+\epsilon)$.
Expanding the result in $\epsilon$, and noting that
(\ref{eq:Ipm0}) and (\ref{eq:Ipminfty}) give
\begin {equation}
   \sum_{m,n=1}^\infty
   \frac{(\pm)^{m+n-1}}{(m+n)^3}
   =
   \zeta_\pm(2) - \zeta_\pm(3)
   = \Delta I_\pm(0) ,
\end {equation}
we get
\begin {equation}
   \int \frac{d^2{\cal Q}}{(2\pi)^2} \>
   \frac{\Delta I_\pm({\cal Q})}{{\cal Q}^{2(1-\epsilon)}}
   =
   \frac{\Delta I_\pm(0)}{4\pi}\
   \left(
      \frac{1}{\epsilon} + 1 - 2\gammaE + 2 \ln 2
   \right)
   - \frac{\sigma_\pm}{2\pi} ,
\label {eq:int2}
\end {equation}
with $\sigma_\pm$ defined as in (\ref{eq:Spm2}).

Evaluating ${\cal I}_\pm = {\cal I}_\pm^{(1)} + {\cal I}_\pm^{(2)}$
of (\ref{eq:qhatdome}) by combining (\ref{eq:Ipm2}), (\ref{eq:int1}),
(\ref{eq:int2}) and
\begin {equation}
   {\cal I}_\pm^{(1)} \equiv
   \int_{q_\perp<\lambda} \frac{d^2 q_\perp}{(2\pi)^2} \>
   \frac{I_\pm(0)}{(q_\perp^2+\md^2)}
   =
   \frac{I_\pm(0)}{2\pi} \ln\left(\frac{\lambda}{\md}\right)
\end {equation}
then produces the final result (\ref{eq:qhateqs}).

% =========================================================================

\section{The NLL calculation}
\label{app:NLL}

For simplicity of presentation, we will focus just on bremsstrahlung
in this appendix.
Arnold and Dogan \cite{ArnoldDogan} computed the NLL bremsstrahlung
rate using the $q_\perp \ll T$ limit in
(\ref{eq:dsig}) for
$d\Gamma_{\rm el}/d^2q_\perp$.  In their notation, they referred to
${\cal A}(q_\perp)$ instead of $d\Gamma_{\rm el}/d^2q_\perp$, and
the translation%
\footnote{
  See Appendix A of Ref.\ \cite{timelpm1}.
}
is
\begin {equation}
   \frac{d\Gamma_{\rm el}}{d^2q_\perp}
   = \frac{C_R g^2}{(2\pi)^2} \, {\cal A}(q_\perp) ,
\end {equation}
\begin {equation}
   \qhat = 
   g^2 \int \frac{d^2q_\perp}{(2\pi)^2} \> {\cal A}(q_\perp) \,
   q_\perp^2 .
\end {equation}
The derivation of Arnold and Dogan is fairly easy to generalize if we
just make these replacements.
If one is confident enough, one can just write our
generalization (\ref{eq:mu}) of Arnold and Dogan by inspection by
recasting their result in terms of the $\Lambda \ll T$ version of
$\qhat$ and then assuming the formula works for $\Lambda \gg T$.
This works because
Arnold and Dogan's constant $\xi$ was generated by the
large-$q_\perp$ contributions to the calculation, where
${\cal A}(q_\perp)$ is proportional to $1/q_\perp^4$ in either case.
It is the large $q_\perp$ part of the calculation that is affected
by the NLL calculation: the $q_\perp \ll Q_\perp$ pieces just come
from the leading-order calculation, which is proportional to
$\qhat(\QTo)$.

Readers may find the above argument obscure, or an explicit
calculation reassuring, and so
we will also indicate how to get the same result by modifying the
calculations of Ref.\ \cite{ArnoldDogan}.
We will not reproduce the entire derivation of Ref.\ \cite{ArnoldDogan}
but will just indicate which equations are modified.

Eq. (4.14) of Ref.\ \cite{ArnoldDogan} for $H_s^2$, which determines
the leading log result, becomes
\begin {equation}
  H^2 =
  \left\{
     2g^2 |p'kp|
     \left[
        \tfrac12\ca p'^2
        +(\cs -\tfrac{1}{2}\ca)k^2
        +\tfrac12\ca p^2
     \right]
     \qhat(Q_{\perp 0})
  \right\}^{1/2} ,
\label {eq:H}
\end{equation}
where $p'=E$ is the initial high-energy parton energy and
$k=x E$ and $p=(1-x)E$ are the energies of the two partons it
splits into.
Combining this formula for $H$ with Eqs.\ (1.1), (4.1), and (4.15) of Ref.\
\cite{ArnoldDogan} gives Eqs.\ (\ref{eq:brem}) and (\ref{eq:LLmu})
of the current paper.  Roughly speaking, $H$ gives the scale of the
total momentum transfer to a high-energy particle during the
formation time as
\begin {equation}
   Q_\perp \sim \frac{H}{x_i E} ,
\end {equation}
where $x_i = 1$, $x$, or $1-x$ depending on which particle one is
focusing on.

In evaluating the NLL correction, the only remaining step that is
different is the evaluation of the integral
\begin {equation}
  I_2(\kappa^2) \equiv
  g^2 \int \frac{d^2h}{(2\pi)^2} \> \frac{d^2 q_\perp}{(2 \pi)^2} \,
    {\cal A}(q_\perp) \,
    {\bm F}_0(\h) \cdot
    \left[ {\bm F}_0(\h)-{\bm F}_0(\h+\kappa \qperp) \right]
\end {equation}
defined in Eq.\ (4.27) of Ref.\ \cite{ArnoldDogan},
with
\begin {equation}
  {\bm F}_0(\h) =
  i \, 4p'kp
  \left[ \exp\left(-e^{i\pi/4} \frac{h^2}{H^2}\right) -1 \right]
  \frac{\h}{h^2}
\end{equation}
and $\kappa = p'$, $p$, or $k$.
Here also, Arnold and Dogan used the small-$q_\perp$ formula for
${\cal A}(q_\perp)$, and we now want to generalize.
Introduce a cut-off $\Lambda \ll Q_\perp$
such that the differential elastic
scattering rate
behaves like
\begin {equation}
   \frac{d\Gamma_{\rm el}}{d^2 q_\perp} \equiv
   C_R \, \frac{d\bar\Gamma_{\rm el}}{d^2 q_\perp}
   \simeq
   C_R \, \frac{c}{(2\pi)^2 q_\perp^4}
   \qquad \mbox{for $\Lambda \le q_\perp \lesssim Q_\perp$}
\label {eq:asymptotic}
\end {equation}
for some constant $c$, up to corrections higher-order in $g$.
We won't need the specific value, but, referring to (\ref{eq:dsig}),
$c=g^4 {\cal N}$ for the case $Q_\perp \gg T$ of interest in this
paper, or $c=g^2 T \md^2$ for the case $\md \ll Q_\perp \ll T$ analyzed
previously in Ref.\ \cite{ArnoldDogan}. 
Now rewrite
\begin {equation}
  I_2(\kappa^2) \simeq
  I_{2<}(\kappa^2) + I_{2>}(\kappa^2) ,
\end {equation}
with
\begin {align}
  I_{2<}(\kappa^2) &\equiv
  \int \frac{d^2h}{(2\pi)^2} \int_{q_\perp<\Lambda} d^2 q_\perp \,
    \frac{d\bar\Gamma_{\rm el}(q_\perp)}{d^2q_\perp} \,
    {\bm F}_0(\h) \cdot
    \left[ {\bm F}_0(\h)-{\bm F}_0(\h+\kappa \qperp) \right] ,
\label {eq:I2l}
\\
  I_{2>}(\kappa^2) &\equiv
  \int \frac{d^2h}{(2\pi)^2}
  \int_{\Lambda<q_\perp} \frac{d^2 q_\perp}{(2\pi)^2} \>
    \frac{c}{q_\perp^4} \,
    {\bm F}_0(\h) \cdot
    \left[ {\bm F}_0(\h)-{\bm F}_0(\h+\kappa \qperp) \right] .
\end {align}
In the integration (\ref{eq:I2l}) for $I_{2<}$, we have
$\kappa q_\perp \ll \kappa Q_\perp \sim H$,
and so one can expand the difference of $F_0$'s in a Taylor expansion:
\begin {align}
  I_{2<}(\kappa^2) &\simeq
  - \frac{\kappa^2}{4}
  \int \frac{d^2h}{(2\pi)^2} \int_{q_\perp<\Lambda} d^2 q_\perp \,
    \frac{d\bar\Gamma_{\rm el}(q_\perp)}{d^2q_\perp} \, q_\perp^2
    {\bm F}_0(\h) \cdot \nabla_h^2 {\bm F}_0(\h)
\nonumber\\
  &=
  - \frac{\kappa^2}{4} \, \qhat(\Lambda)
  \int \frac{d^2h}{(2\pi)^2}
    {\bm F}_0(\h) \cdot \nabla_h^2 {\bm F}_0(\h)
\nonumber\\
  &=
  - \frac{2}{\pi} \, \kappa^2 \, \qhat(\Lambda) \,
  (p'kp)^2 \frac{e^{i\pi/4}}{H^2}
  \,.
\end {align}
Now turn to $I_{2>}$.  It is convenient to rewrite
\begin {equation}
  I_{2>} =
  I_{2>}^{\rm a}
  + I_{2>}^{\rm b}
\end {equation}
with
\begin {align}
  I_{2>}^{\rm a} &=
  \int \frac{d^2h}{(2\pi)^2} \int \frac{d^2 q_\perp}{(2\pi)^2} \,
    \left(\frac{c}{q_\perp^4} \, \theta(q_\perp-\Lambda)
        - \frac{c}{q_\perp^2(q_\perp^2+M^2)}\right) \,
    {\bm F}_0(\h) \cdot
    \left[ {\bm F}_0(\h)-{\bm F}_0(\h+\kappa \qperp) \right] ,
\\
  I_{2>}^{\rm b} &=
  \int \frac{d^2h}{(2\pi)^2} \int \frac{d^2 q_\perp}{(2\pi)^2} \,
    \frac{c}{q_\perp^2(q_\perp^2+M^2)} \,
    {\bm F}_0(\h) \cdot
    \left[ {\bm F}_0(\h)-{\bm F}_0(\h+\kappa \qperp) \right] .
\end {align}
Here $\theta(z)$ is the step function and
$M \lesssim \Lambda$ is an arbitrary scale.
In the first integral, we can again treat $\kappa q_\perp$ as small
and Taylor expand the difference in $F_0$'s, giving
\begin {align}
  I_{2>}^{\rm a} &\simeq
  - \frac{\kappa^2}{4}
  \int \frac{d^2 q_\perp}{(2\pi)^2}
  \left(\frac{c}{q_\perp^4} \, \theta(q_\perp-\Lambda)
        - \frac{c}{q_\perp^2(q_\perp^2+M^2)}\right) \,
  q_\perp^2
  \int \frac{d^2h}{(2\pi)^2}
  {\bm F}_0(\h) \cdot \nabla_h^2 {\bm F}_0(\h)
\nonumber\\
  &= - \frac{\kappa^2}{4}
  \times \frac{c}{2\pi} \, \ln\left(\frac{M}{\Lambda}\right)
  \times \frac{8}{\pi} (p'kp)^2 \frac{e^{i\pi/4}}{H^2} \,.
\end {align}
$I_{2>}^{\rm b}$ is proportional to the $I_2(\kappa^2)$ integral
evaluated by Arnold and Dogan, with $\md^2$ replace by $M^2$ in the
denominator, and the overall normalization $g^2 T \md^2$
replaced by $c$.  So we can take over the result from
Eqs. (4.31) and (4.32) of Ref.\ \cite{ArnoldDogan},
\begin {equation}
  I_{2>}^{\rm b} \simeq
  - \frac{(p'kp)^2 c}{\pi^2 M^2} \,
  ( 2-\gammaE-\ln u_\kappa ) u_\kappa
\end {equation}
with
\begin{equation}
   u_\kappa \equiv e^{i\pi /4}\, \frac{M^2 \kappa^2}{2 H^2} \,.
\end{equation}
Combining the various pieces above,
\begin {equation}
  I_2(\kappa^2) \simeq
  - \frac{2 \kappa^2 e^{i\pi/4} (p'kp)^2}{\pi H^2} \,
  \left\{
    \qhat(\Lambda)
    -\frac{c}{4\pi} \left[
        -2 + \gammaE
        + \ln\left( e^{i\pi/4}\,\frac{\kappa^2\Lambda^2}{2 H^2}\right)
    \right]
  \right\} ,
\end {equation}
which replaces Eq.\ (4.32) of Ref.\ \cite{ArnoldDogan}.
Now take the real part of $I_2$ and note that
(\ref{eq:qhatdef}) and (\ref{eq:asymptotic}) implies that
\begin {equation}
   \qhat(\Lambda')
   \simeq \qhat(\Lambda)
   + \frac{c}{2\pi} \, \ln\left( \frac{\Lambda'}{\Lambda} \right)
\end {equation}
for $\Lambda'$ and $\Lambda$ both in the region covered by
(\ref{eq:asymptotic}).
Using the definition (\ref{eq:xi}) of $\xi$, we then get
\begin {equation}
  \Real I_2(\kappa^2) \simeq
  - \frac{\sqrt2 \kappa^2 (p'kp)^2}{\pi H^2} \,
  \qhat\biggl(\sqrt{\frac{2\xi H^2}{\kappa^2}}\biggr)
  ,
\end {equation}
which replaces Eq.\ (4.33) of Ref.\ \cite{ArnoldDogan}.
Eq.\ (4.35) of that reference then becomes
\begin {align}
  & \Real({\bm S},{\bm F}_s)
  =
  \Real({\bm S},{\bm F}_0) \,
\nonumber\\ & \qquad \times
  \frac12
  \left\{
    1 +
    \frac{
       \tfrac12\ca p'^2 \,
            \qhat\left(\sqrt{\frac{2 \xi H_s^2}{p'^2}}\right)
       +(\cs -\tfrac{1}{2}\ca)k^2 \,
            \qhat\left(\sqrt{\frac{2 \xi H_s^2}{k^2}}\right)
       +\tfrac12\ca p^2 \,
            \qhat\left(\sqrt{\frac{2 \xi H_s^2}{p^2}}\right)
    }{
       \tfrac12\ca p'^2 \, \qhat(Q_{\perp 0})
       +(\cs -\tfrac{1}{2}\ca)k^2 \, \qhat(Q_{\perp 0})
       +\tfrac12\ca p^2 \, \qhat(Q_{\perp 0})
    }
  \right\} ,
\end {align}
where the notation $({\bm S},{\bm F}_s)$
is defined in Ref.\ \cite{ArnoldDogan}.
Following the same steps as Arnold and Dogan then produces
our result (\ref{eq:mu}), where our $\mu_\perp$ corresponds to their
$\md \hat\mu_\perp$.

%%%%%%%%%%%%%%%%%%%%%%%%%%%%%%%%%%%%%%%%%%%%%%%%%%%%%%%%%%%%%%%%%%%%%%%%%%%%%%

\end{document}